\documentclass{PoS}

\usepackage{amsmath,subfigure,amsfonts,latexsym}
\usepackage{amssymb}
\usepackage{epsfig}
\usepackage{graphicx}

\title{Light neutralino dark matter in light Higgs scenario related with the CoGeNT and DAMA/LIBRA results}

\ShortTitle{Light neutralino dark matter in light Higgs scenario}

\author{\speaker{Seodong Shin}\\
        FPRD and Department of Physics and Astronomy, 
        Seoul National University, \\
        Seoul 151-747, Korea \\
        E-mail: \email{sshin@phya.snu.ac.kr}}

\abstract{Recently, the CoGeNT collaboration reported the WIMP candidate signal events exceeding the known backgrounds where the light WIMP with large cross section is supported. Motivated by this issue, we analyze a light neutralino dark matter scenario with a very light CP-even Higgs mediation in the elastic scattering process, which provides the mass and direct detection cross section to explain the CoGeNT result. To be compatible with the result of other experiments such as LEP and B-factories, the light CP-even Higgs is favored to be in 9 to 10 GeV. Such a scenario can be realized in the ``Beyond the MSSM" context. The relic abundance consistent with the WMAP result can be obtained when twice of neutralino mass is close to the light Higgs mass via the resonance enhancement of the annihilation cross section. As a result, the neutralino mass is predicted to be at around 5 to 6 GeV.
}

\FullConference{Identification of Dark Matter 2010-IDM2010\\
		July 26-30, 2010\\
		Montpellier France}

\begin{document}

\section{Introduction}

Various candidates of CDM explaining the observed relic abundance $\Omega h^2 \sim 0.1$ has been proposed. Among them, the most promising one is Weakly Interacting Massive Particle (WIMP) such as the lightest supersymmetric particle (LSP) in the supersymmetric models with R-parity and the SM gauge singlet particles in Higgs portal models \cite{SDM}. WIMPs are produced at the early stage of our universe and their current relic abundance is naturally determined when their interactions to the SM particles freeze out \cite{Lee:1977ua}.

Recently, the Coherent Germanium Neutrino Technology (CoGeNT) experiment reported that about a hundred events \cite{cogent} exceeding the expected background are observed after their eight week operation, which possibly originated from the nuclear recoil by DM scattering. Due to its enhanced sensitivity to low energy events, the ionization signal observed in the CoGeNT detector is as low as $0.4 - 3.2$ keVee. The discovery region, hence, supports the existence of light DM whose mass is $m_{\chi} \lesssim 10$ GeV and the spin independent WIMP-nucleon elastic scattering cross section $\sigma_{\rm SI}$ is as high as $\sim 10^{-40}$ cm$^2$. The light WIMP with large cross section is also favored in the recent DAMA/LIBRA annual modulation signal \cite{dama}. Considering the channeling effect \cite{channeling}, the result allows sizable parameter space compatible with other null experiments. There have been several proposals to explain this, such as light SM gauge singlet fermion \cite{SFDMdama}. Therefore, it is very interesting to propose a plausible light WIMP scenario\footnote{It must be noted that there is a negative research such that the channeling fraction of recoiling lattice nuclei in NaI is quite suppressed to provide its meaningful effect \cite{Bozorgnia:2010xy}.}. 

There have been researches to reconcile the CoGeNT report on WIMP signals with other null experiments constraining the detection bound and the previous DAMA result \cite{cogentpapers1,cogentpapers2,Bae:2010hr}. In this proceeding, we briefly summarize the analysis of the light neutralino in light Higgs scenario \cite{Bae:2010hr}. We focus on the realization of light neutralino of mass $4 - 7$ GeV, which is not so much constrained by recent analysis of XENON100 \cite{xenon100} and XENON10 \cite{Sorensen:2010hq}.

\section{Light neutralino}

 In order to have a large scattering cross section of the neutralino off the target nuclei in a supersymmetric SM, either large value of $\tan\beta$ or a very light CP-even Higgs boson mediator is needed in the process of elastic scattering of the neutralino off the target nuclei. Such scenarios cannot be easily realized in the context of the MSSM since they are highly constrained by other experiments such as the LEP, Tevatron, and rare decays. Especially, it is not easy to obtain the $\sigma_{\rm SI}$ as high as $\sim 10^{-40}$ cm$^2$ with conventional halo parameters. Therefore, the "Beyond the MSSM" (BMSSM) considerations are required here. Without considering additional light degrees of freedom, the light Higgs scenario is natural to be analyzed. (Scenarios with light degrees of freedom are explained in \cite{lightnmssm} with the NMSSM context and in \cite{SFDMdama} with the SM gauge singlet Dirac fermion. The latter can be also obtained by slightly changing the NMSSM potential with singlet quadratic terms.)

\subsection{Light neutralino in the heavy Higgs scenario}

In the regime of very large $\tan\beta$ with $m_H$ not much larger than $m_h$, neutralino-nucleon scattering process is dominated by the heavy CP-even Higgs mediated contribution so that we obtain 
\begin{equation}
\sigma_{\rm SI} \simeq 0.23 \times 10^{-40}\mbox{ cm}^2\times\biggl(\frac{N_{13}}{0.4}\biggr)^2
\biggl(\frac{\tan\beta}{50}\biggr)^2\biggl(\frac{100~{\rm GeV}}{m_H}\biggr)^4~, \label{eq:detect}
\end{equation}
for $m_{\chi} \sim 7$ GeV where the subdominant down quark and one-loop induced bottom quark contributions are also considered. Therefore, we need very large $\tan\beta > 100$ for $m_H=100$ GeV to explain the CoGeNT result. In the regime of such large $\tan\beta$, however, the branching ratio of $B_s \to \mu^{+}\mu^{-}$ severely constrains the parameter set. In addition, combining the the upper limit on the neutral Higgs bosons $\to \tau^{+} \tau^{-}$ in the Tevatron and the observations of the rare decays $B \to \tau \nu$ with the ratio of $B \to D \tau \nu / B \to D l \nu$ in the B factories, we obtain the constraints on the elastic scattering $\sigma_{\rm SI} \lesssim 5 \times 10^{-42}$ cm$^2$, which is much lower than $10^{-40}$ cm$^2$ to explain the CoGeNT result \cite{Feldman:2010ke,Kuflik:2010ah}. In the heavy $H$ scenario, it is hence very hard to construct viable models which support the light WIMP of mass $4 - 7$ GeV with the spin independent elastic cross section $\sigma_{\rm SI} \sim 10^{-40}\mbox{ cm}^2$. In order to reduce low energy constraints, we need to invoke ``wrong-Higgs" interactions \cite{Bae:2010ai}.

\subsection{Light neutralino in the light Higgs scenario}

Instead of the heavy CP-even Higgs mediation scenario for very large $\tan\beta$, we consider the case that the light CP-even Higgs mediated contribution dominates the elastic scattering with moderate $\tan\beta\simeq3$. The light Higgs mediation can be more important than that of heavy Higgs if $m_h \ll m_H$. In such a case, light Higgs mass $m_h$ is very small to have large cross section for $\sigma_{\rm SI}$ so that constraints from LEP experiments and rare decays of mesons must be considered. 

If $10$ GeV$\lesssim m_h\lesssim20$ GeV, two kinds of Higgs search at the LEP experiments must be under consideration. One is the Higgsstrahlung process, $e^+e^-\to Z^*\to hZ$,
and the other is the associative production, $e^+e^-\to Z^* \to hA$. For convenience, we define the following quantities
\begin{eqnarray}
R_{hZ}&\equiv&
\frac{\sigma(e^+e^-\to Z^*\to Zh)_{\rm MSSM}}{\sigma(e^+e^-\to Z^*\to Zh)_{\rm SM}} 
=\sin^2(\alpha-\beta) , \\
& & \nonumber \\
R_{hA}&\equiv&\frac{\sigma(e^+e^-\to Z^*\to hA)_{\rm MSSM}{\cal B}(h\to \bar{b}b){\cal B}(A\to \bar{b}b)}
{\sigma(e^+e^-\to Z^*\to hA)_{\rm ref}}\\
&=&\cos^2(\alpha-\beta){\cal B}(h\to \bar{b}b){\cal B}(A\to \bar{b}b) ,\nonumber
\end{eqnarray}
where $\sigma(e^+e^-\to Z^*\to hA)_{\rm ref}$ is a reference value assuming that 
$Z-h-A$ coupling constant is equal to that of the SM $Z-Z-h$ coupling, {\it i.e.}, $g_{ZhA}=g_{ZZh}^{\rm SM}$. The value of $\alpha$ is the Higgs mixing angle. 
In order to satisfy the negative results of scalar searches at the LEP experiments, $R_{hZ} \lesssim 0.01$ is required for the Higgsstrahlung process and $R_{hA} \lesssim 0.2$ for the associative production when $m_h\sim 20$ GeV, $m_A\sim 90$ GeV \cite{LHWG-Note:2005-01}.

In the case that $\sin^2(\alpha-\beta)<0.01$, we can evade the constraint from the Higgsstrahlung process.
However, avoiding the associative production constraint is not trivial since $\cos(\alpha-\beta)\simeq1$.
The light neutralino with $m_{\chi}\lesssim m_h /2$ can be a solution in this case. Since the light Higgs can decay to the neutralinos, the branching ratio of Higgs decay to neutralinos is comparable to or larger than that of Higgs decay to $b$-quark pair for low $\tan\beta\lesssim3$ \cite{Yaguna:2007vm}. Consequently, we have reliable parameter space by constraining $m_{h} \gtrsim 2 m_{\chi}$ when $10$ GeV $\lesssim m_h \lesssim$ $20$ GeV. It is, however, not the end of the story and this region will be discussed again in the next section.

If $m_h<10$ GeV, light Higgs cannot decay to $b$-quark pair but can decay to $\tau$-leptons.
In this case, the constraint from the associative production is practically not relevant
because Higgs constraints from $2b2\tau$ final state is much weaker than those from $4b$ final state \cite{LHWG-Note:2005-01}.
On the other hand, the constraint from radiative $\Upsilon$ decay, 
$\Upsilon\to h\gamma$ is on rise as well as Higgsstrahlung constraint $\sin(\alpha-\beta) \approx 0$. 
Very light scalar particles can contribute the radiative decay of $\Upsilon$-meson \cite{Wilczek:1977zn}. Taking the strongest bound, ${\cal B}(\Upsilon\to h\gamma)<10^{-5}$ \cite{:2008hs,Aubert:2009cka},
we obtain a conservative bound 
\begin{equation}
m_h^2>m_{\Upsilon}^2\biggl[0.894-0.0150\biggl(\frac{\tan\beta}{3}\biggr)^{-2}\biggr].
\label{bound:upsilon}
\end{equation}
Hence, we obtain $m_h\gtrsim8.9$ GeV for $\tan \beta=3$ to evade the radiative $\Upsilon$ decay constraints.

Under such considerations, neutralino-nucleon scattering cross section 
is obtained in the case of $\sin^2(\alpha-\beta)<0.01$, {\it i.e.}, $\alpha\simeq\beta$, such that
\begin{eqnarray}
\sigma &\simeq& 4.7\times10^{-40}{\rm cm}^2\times\biggl(\frac{N_{13}}{0.3}\biggr)^2
\biggl(\frac{\tan\beta}{3}\biggr)^2\biggl(\frac{10~{\rm GeV}}{m_h}\biggr)^4,
\label{sigma:h}
\end{eqnarray}
for $\tan \beta \sim \mathcal{O}(1)$, as we have expected. Such a light Higgs scenario is not obtained in ordinary MSSM parameter space.
Instead, we consider BMSSM \cite{Dine:2007xi,Kim:2009sy,Bae:2010cd} where
light Higgs scenarios can be realized \cite{Bae:2010cd}.
The $SU(2)$ doublet CP-even Higgs mass matrix in the basis of $(H_d^0, H_u^0)$ is 
\begin{equation}
\begin{pmatrix}
M_Z^2c^2_{\beta}+m_A^2s^2_{\beta}-4v^2\epsilon_1s_{2\beta}+4v^2\epsilon_2s^2_{\beta}&
-(M_Z^2+m_A^2)s_{\beta}c_{\beta}-4v^2\epsilon_1\\
-(M_Z^2+m_A^2)s_{\beta}c_{\beta}-4v^2\epsilon_1 &
M_Z^2s^2_{\beta}+m_A^2c^2_{\beta}-4v^2\epsilon_1s_{2\beta}+4v^2\epsilon_2c^2_{\beta}
\end{pmatrix}\label{Higgs:mass}
\end{equation}
where $M_Z$ is the mass of $Z$ boson, $m_A$ is the mass of the CP-odd Higgs,  $s_{\beta}$($c_{\beta}$) is $\sin\beta$($\cos\beta$), 
and $\epsilon_{1,2}$ are BMSSM parameters defined by \cite{Dine:2007xi}. Here, $\epsilon_{1,2}$ are assumed to be real for simplicity. 

The physical CP-even Higgs bosons are obtained as the mass eigenstates of (\ref{Higgs:mass}),
\begin{equation}
\begin{pmatrix}
H\\h
\end{pmatrix}
=
\begin{pmatrix}
\cos\alpha & \sin\alpha \\ -\sin\alpha & \cos\alpha
\end{pmatrix}
\begin{pmatrix}
H_d^0 \\ H_u^0
\end{pmatrix} \label{eq:pmatrix}
\end{equation}
where $m_h<m_H$ and $-\pi/2 \le \alpha \le \pi/2$ in contrast to the MSSM case.
In the MSSM case where $\epsilon_{1,2}=0$ and $\tan\beta\lesssim 5$, $m_A$ cannot be larger than $M_Z$ since $h$ must be light enough to obtain the large direct detection cross section such as (\ref{sigma:h}). If $m_A\ll M_Z$, light Higgs $h$
is mostly down-type and $\alpha\simeq\pi/2$ so that the constraint from the Higgsstrahlung process $Z^*\to hZ$ can be evaded. But the associative production $e^+e^-\to hA$ still constrains such a case. As previously discussed, it seems possible to avoid this constraint if Higgs bosons decay to neutralinos, however, there still remain other obstacles. If $m_h+m_A<M_Z$, the invisible decay width of $Z$-boson must be considered, hence, such parameter region cannot be viable. 
On the other hand, if $m_A\lesssim M_Z$, Higgs mixing is maximized so that $m_h\ll M_Z$ and
$m_H$ is larger than LEP search bound 114 GeV.
In this case, however, the Higgsstrahlung constraint for light Higgs can not be avoided since $\alpha\sim -\pi/4$. Therefore, it seems that very light CP-even Higgs scenario is not realized in the MSSM context.

Considering the BMSSM, instead, the analysis on CP-even Higgs mass is quite different due to the existence of non-zero $\epsilon_{1,2}$.
Observing mass matrix (\ref{Higgs:mass}), the negative $\epsilon_2$ correction effectively reduces $m_A^2$ in the MSSM to $m_A^2+4v^2\epsilon_2$ so that very light $h$ scenario can be realized   
without introducing light CP-odd Higgs. In addition, negative $\epsilon_1$ correction can reduce the off-diagonal Higgs mixing. Moreover, when $4 v^2 |\epsilon_1| \gtrsim (M_Z^2 + m_A^2) s_{\beta} c_{\beta}$, we achieve $\alpha \lesssim \pi/2$ so that the condition $\sin^2 (\beta - \alpha) < 0.01$ is satisfied. Therefore, it seems plausible to obtain a light CP-even Higgs scenario in the context of BMSSM.\footnote{The issue on the stability of the BMSSM Higgs potential will be discussed in the revised version of \cite{Bae:2010hr}.}

\section{Numerical results in BMSSM parameter space}

\begin{figure}
\begin{center}
\subfigure[\ $\tan\beta=3$, $m_A=90$ GeV]{
\includegraphics[width=6cm]{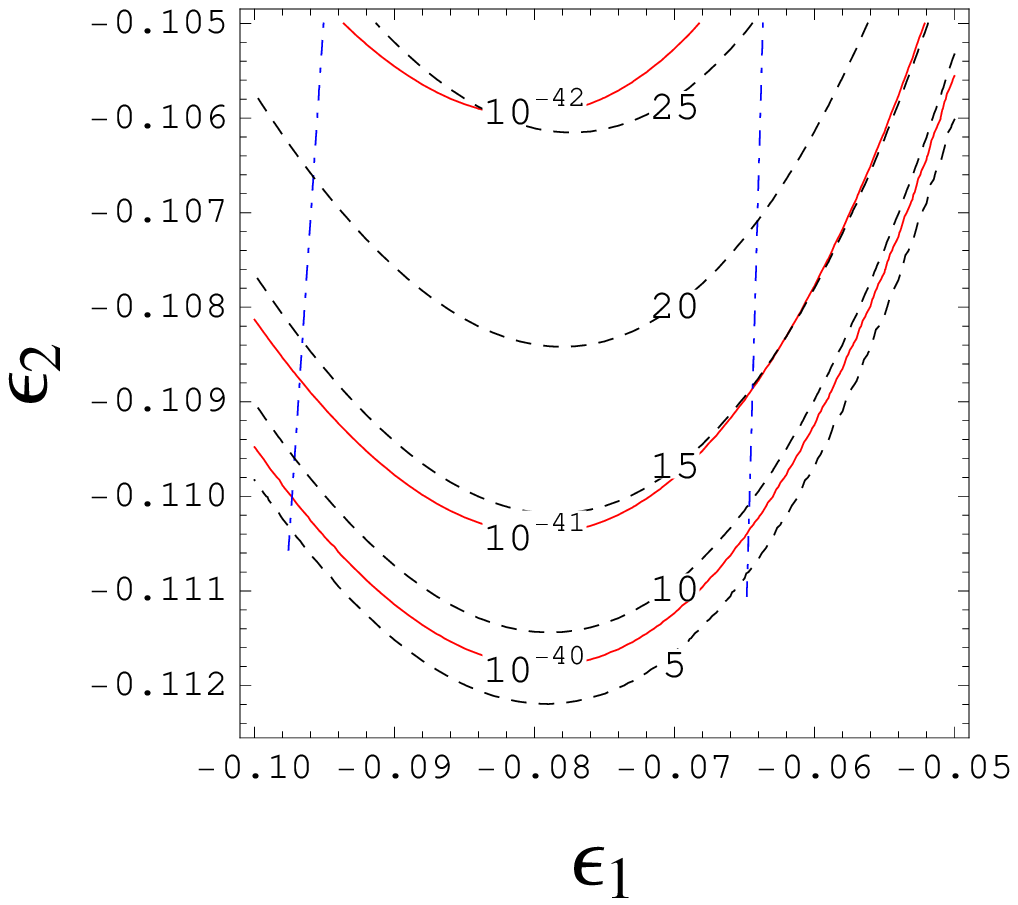}}
\quad
\subfigure[\ $\tan\beta=5$, $m_A=90$ GeV]{
\includegraphics[width=6cm]{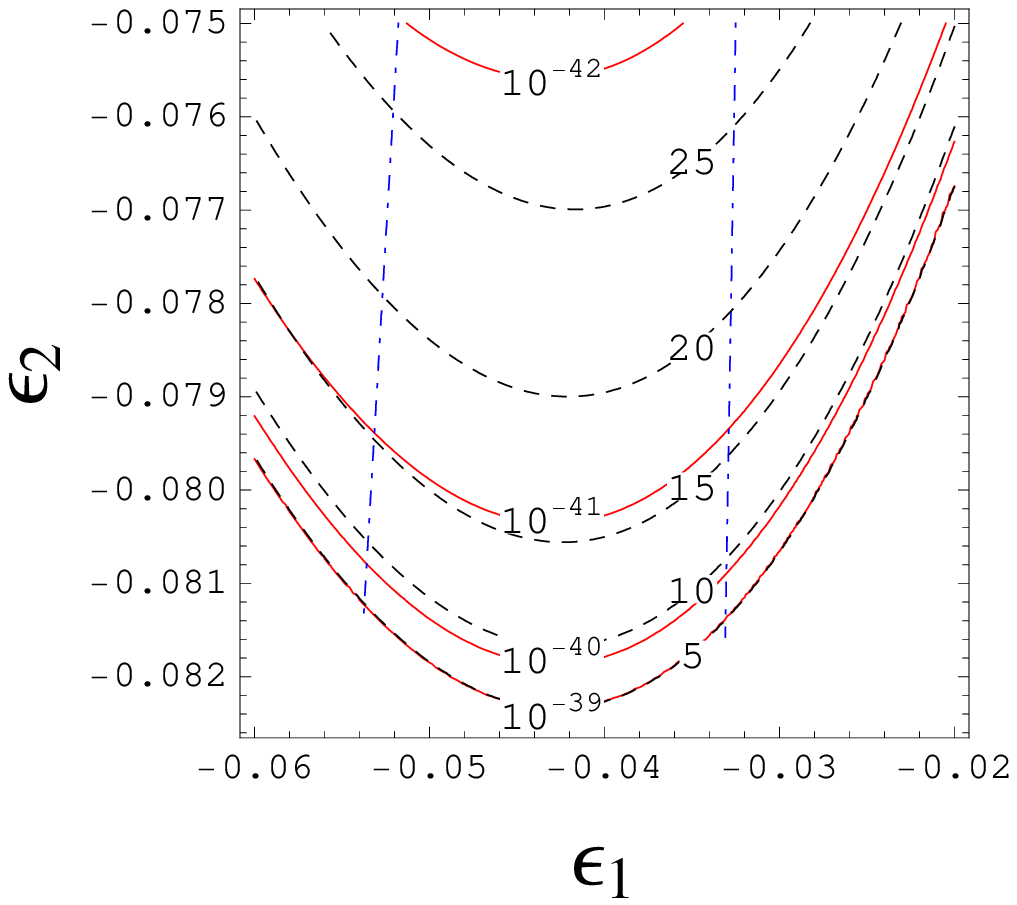}}
\end{center}
\caption{Numerical results of neutralino-nucleon scattering cross section in 
$\epsilon_1-\epsilon_2$ plane. Solid(red) curves stand for scattering cross section,
$\sigma=10^{-40}$-$10^{-42}$ cm$^2$ for left panel and $\sigma=10^{-39}$-$10^{-42}$ cm$^2$ for right panel
from the bottom to the top.
Dashed(black) curves stand for light Higgs mass, $m_h=5$-$25$ GeV 
from the bottom to the top. Dot-dashed(blue) lines stand for $\sin^2(\alpha-\beta)=0.01$,
the region between the lines is safe from the Higgsstrahlung constraint.
Neutralino mass is $4$ GeV$\lesssim m_{\chi} \lesssim7$ GeV depending on parameters.}
\label{fig:result}
\end{figure}

In the BMSSM, the $\epsilon_1$ correction is also included in the neutralino sector \cite{Dine:2007xi,Berg:2009mq}. The coupling $g_{\phi\chi\chi}$ is modified by $\epsilon_1$ term such that \cite{Berg:2009mq}
\begin{eqnarray}
\delta g_{h\chi\chi} &=& -\frac{2\epsilon_1}{\mu}\big(v\sqrt{2}\cos\beta\cos\alpha (N_{14})^2
+v\sqrt{2}\sin\beta\sin\alpha (N_{13})^2\nonumber\\
&&+2\sqrt{2}v\sin(\alpha+\beta)N_{13}N_{14}\big),\label{feyn:e1_hNN}\\
\delta g_{A\chi\chi} &=& -\frac{2\epsilon_1}{\mu}\big(iv\frac{1}{\sqrt{2}}\sin2\beta(N_{14})^2
+iv\frac{1}{\sqrt{2}}\sin2\beta(N_{13})^2+i2\sqrt{2}vN_{13}N_{14}\big),\label{feyn:e1_ANN}
\end{eqnarray}
where $\mu$ is the Higgsino mass parameter in the MSSM superpotential. Since the BMSSM corrections are much smaller than the MSSM ones, these corrections do not spoil the aforementioned advantages. Numerical results are given in Fig. \ref{fig:result}. From the figures, $-0.10\lesssim\epsilon_1\lesssim -0.06$, $\epsilon_2\sim-0.11$ for $\tan\beta=3$
and $-0.06 \lesssim\epsilon_1\lesssim -0.03$, $\epsilon_2\sim-0.08$ for $\tan\beta=5$
give the desired scattering cross section for CoGeNT results, simultaneously  satisfying the LEP Higgsstrahlung constraint. 

In the case of $m_h>10$ GeV, however, this parameter space is not allowed due to the large value of $R_{hA}\sim0.3>0.2$. We, hence, need lower value of $\tan \beta$ since the branching ratio of $h$ to $\chi \chi$ becomes larger for smaller $\tan \beta$. Comparing the Fig. \ref{fig:result}(a) and Fig. \ref{fig:result}(b) , however, larger values of $|\epsilon_{1,2}|$ are needed to obtain the light Higgs spectrum for smaller $\tan \beta$. Then, we need much fine tuning to obtain such light Higgs so that it is more natural to consider only the case $m_h<10$ GeV. Consequently, the light Higgs scenario to explain the light neutralino of $4$ GeV $\lesssim m_{\chi} \lesssim$ $7$ GeV and $\sigma_{\rm SI} \sim 10^{-40}$ cm$^2$ is most naturally realized when $\sin^2(\alpha-\beta)<0.01$ and $9$ GeV$\lesssim m_h\lesssim10$ GeV.

\section{Relic Abundance}

\begin{figure}
\begin{center}
\includegraphics[width=7.7cm]{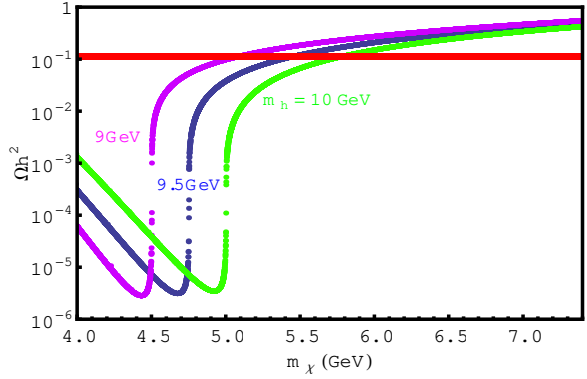}
\end{center}
\caption{$\Omega h^2$ to $m_{\chi}$ with fixed $\tan \beta =3$ and $N_{13} = 0.3$ values. The allowed mass of the light CP-even Higgs is 9 GeV $\lesssim m_h \lesssim$ 10 GeV. The magenta, blue, green lines denote the case $m_h = 9$ GeV, 9.5 GeV, 10 GeV respectively. The red region denotes the observed relic abundance.}
\label{fig:relic}
\end{figure}

Since the neutralino is very light, $m_{\chi}\lesssim7$ GeV, they annihilate only to light fermions at the freeze-out. In addition, the neutralino is much lighter than the CP-odd Higgs $A$, squarks, and $Z$-boson so that the dominant annihilation process is mediated by the CP-even Higgs $h$ which is a P-wave process. Furthermore, small $\tan\beta \sim 3$ constrains the interaction of $h$ to the SM fermions. Therefore, one might worry that the neutralino will overclose the universe. There is one way out, however. Since the mass of the light CP-even Higgs is highly constrained, $9~{\rm GeV} \lesssim m_h \lesssim 10~{\rm GeV}$, the resonant annihilation of the light WIMPs to the SM fermions through the $s$-channel process can dominate the annihilation process and reduce the relic abundance at the freeze-out. 

We show the role of resonant annihilation to obtain the observed relic density in Fig.\ref{fig:relic}. It is clear that the expected relic abundance decreases around the resonance region so that the mass of the light neutralino is determined within $5$ GeV$ \lesssim m_{\chi} \lesssim$ $6$ GeV. The physically consistent range of $m_{\chi}$ is similar for other possible parameter choice since the right relic abundance will be obtained within the the resonant annihilation process.

\section{Conclusions}

The light WIMPs with large cross section are focussed on due to the recent results of direct detection experiments such as CoGeNT and DAMA. We showed that such a light neutralino can be obtained from the light CP-even Higgs scenario. In the context of the MSSM, however, the existence of such a light Higgs is highly constrained by the LEP experiments. Instead, we looked for a possibility of explaining $\sigma_{\rm SI} \sim 10^{-40}$ cm$^2$, $m_{\chi} \sim$ 4 to 7 GeV dark matter within the framework of the BMSSM (MSSM field contents at and below the weak scale)
and found that $m_h \sim$ 9 to 10 GeV can provide the required $\sigma_{\rm SI}$. If we require the model to explain the right relic abundance, $m_{\chi}$ is predicted to be in between 5 to 6 GeV
depending on the light CP even Higgs mass 9 to 10 GeV.

\end{document}